# High-pressure torsion deformation induced phase transformations and formations: new material combinations and advanced properties

Andrea Bachmaier[*] and Reinhard Pippan

Erich Schmid Institute of Materials Science, Austrian Academy of Sciences, Jahnstrasse 12, 8700 Leoben, Austria

*Corresponding author, E-mail: andrea.bachmaier@oeaw.ac.at

**Abstract**

Heavy plastic shear deformation at relatively low homologous temperatures is called high-pressure torsion (HPT) deformation, which is one method of severe plastic deformation (SPD). The aim of the paper is to give an overview of a new processing approach which permits the generation of innovative metastable materials and novel nanocomposites by HPT deformation. Starting materials can be either coarse-grained multi-phase alloys, a mixture of different elemental powders or any other combination of multiphase solid starting materials. After HPT processing, the achievable microstructures are similar to the ones generated by mechanical alloying. Nevertheless, one advantage of the HPT process is that bulk samples of the different types of metastable materials and nanocomposites are obtained directly during HPT deformation. It will be shown that different material combinations can be selected and materials with tailored properties, or in other words, materials designed for specific applications and the thus required properties, can be synthesized. Areas of application for these new materials range from hydrogen storage to materials resistant to harsh radiation environments.

**Keywords:** severe plastic deformation, high-pressure torsion, phase transformation, mechanical alloying, amorphization, crystallization, metastable phases, nanocomposite, nanostructure

1. Introduction

In 1935, P. Bridgman introduced a processing method, in which disc-shaped samples are heavily shear deformed between two anvils at relatively low homologous temperatures under a high hydrostatic pressure [1]. Bridgman already applied this innovative method to many different kinds of materials and compounds – an extensive review over this early period can be found in ref. [2]. In the late 1980s, the method was named 'high-pressure torsion' (HPT) [3] and rapidly gained ground as a scientific tool to obtain a broad range of ultrafine grained (UFG) or nanocrystalline (NC) metals and alloys with outstanding mechanical properties [4–6]. One major advantage of HPT processed materials is, for example, their ultra-high strength. Even single-phase materials achieve small grain sizes of few 100 nm, which is accompanied by an improvement in mechanical properties keeping all their other beneficial properties.

HPT is, however, not only a well-established technique to achieve grain refinement, but also to induce various phase transformations [7,8]. HPT deformation can cause dissolution of phases [7,8], disordering of ordered phases [9] as well as amorphization of crystalline phases [10] or crystallization of nanocrystals in the amorphous matrices [11]. It further enables the synthesis of metastable phases (e.g. low-temperature, high-temperature or high-pressure allotropic modifications [12–18]). Additionally, phase formations (supersaturated solid solutions) or phase separations during shear deformation are reported [19–23]. Depending on the amount and type of the different mutually immiscible components, either NC supersaturated alloys or nanocomposites with partial supersaturation, not producible by classical metallurgical ways, can be synthesized.

In general, the HPT-processed microstructures are similar to the ones generated by mechanical alloying. The product of HPT is, however, a bulk material with a well-defined composition. The starting materials can be very diverse, like coarse-grained multi-phase alloys, a mixture of powders or any other combination of solid starting materials. By using powders, a wider range of compositions becomes possible since conventional casting is often problematic due to large miscibility gaps in the used systems. Although the HPT processing method has already been successfully applied to different systems, the detailed processes to explain strain-induced mechanical mixing, metastable phase formations, amorphization or crystallization have not been entirely clarified until now. Nevertheless, there is now intense activities in the HPT community in this research field.

In this article, an overview on studies carried out during the past 20 years using the above-mentioned HPT induced phase transformations and formations to synthesize bulk novel NC materials is given. The article further aims to highlight very recent achievements and new trends in this active and developing research field. The overview is divided in the following chapters: First, a short review on deformation-induced phase transformation and formation processes is given. The focus of this overview is, however, on the properties of these novel materials and nanocomposites. Excellent structural properties, for example their high strength and microstructural stability during elevated temperatures, make these materials ideal candidates for fabrication of miniaturized devices, i.e. microelectromechanical systems. Additionally, HPT processing is used as an innovative solid-state route for the synthetization of NC materials with tailored physical properties, i.e. for solid-state hydrogen storage or radiation tolerant behavior. Finally, recent trends of creating new material combinations - bulk metallic glass composites or high-entropy alloys obtained from powders - as well as the use of innovative starting materials - coated or gas atomized powders - for homogeneous HPT induced mechanical alloying and an acceleration of the deformation-induced mixing process, are discussed.

2. Phase transformations and formations during HPT processing

During HPT deformation, a disc shaped sample is put between two anvils, which both have a cylindrical cavity with a depth somewhat smaller than the thickness of the specimen. If powder mixtures are used as starting material, the powders can be either filled directly into the anvils and consolidated in the HPT tool or they can be hydrostatically pre-compacted in air or in inert atmosphere to avoid contamination. During deformation, one of the HPT anvils is fixed, whereas the other one rotates. Typically, a pressure of several GPa is applied during the deformation process. The sample is ideally deformed by simple shear. If this is the case, the various initial phases in a binary or multiple-phase system are elongated in shear direction and their thicknesses are continuously reduced. To obtain a saturation or steady state in a multiphase system, this means that there is no further change in the microstructural features of the sample during on-going HPT deformation, the applied shear strain has to be sufficient to decrease the thicknesses of the sheared phases to the nanometer level. Starting with a multiphase system with initial phase sizes of 50 µm, the applied shear strain should be at least 50,000 to achieve a steady state in the case of ideal co-deformation. If the components are immiscible, nanostructured composites can be synthesized by this way. In reality, ideal shear and co-deformation of phases does hardly ever occur. Very often fragmentation of one phase and localization of the shear deformation occur. If an inhomogeneous deformation takes place, the applied shear strain can be even higher, but under certain conditions it can be smaller too [24].

Furthermore, amorphization processes and the formation of metastable solid solution phases (mechanical alloying) are competing processes if the respective phase sizes reaches the nanometer level. For the observed mechanical alloying processes, different mechanisms involving dislocations or enhanced atomic mobility due to point defects have been proposed [25–29]. In these studies, mechanical alloying is treated as a phenomenon observed in non-equilibrium processing in general and the proposed mechanism are not restricted to occur solely during HPT deformation, but during all other SPD processes, wire drawing and ball milling. In [30], it is proposed that during SPD deformation a thermodynamic driving force for dissolution can be obtained, if the phase size can be decreased to the nanometer level. Mechanical mixing is then achieved by dispersing the dissolved atoms in the alloy matrix. Veltl et al. [31] discussed that the energy stored in the grain boundaries of NC materials might be another possible solid solution formation driving force. In [32], a capillary pressure is suggested as driving force for phase dissolution and mechanical alloying. Next are plasticity-driven mechanism: in the kinetic roughening model of Bellon and Averback [33], atoms are suggested to be shifted across phase boundaries. By the shear of atomic glide planes during deformation, the phase boundaries are increasingly roughened leading to complete chemical mixing of multiphase system. A similar mechanism is discussed for cementite dissolution during deformation of pearlitic steels [34] and in the dislocation shuffle mechanism as proposed in [35]. In [36] it was further shown that the homogeneity of the deformation process strongly influences the degree of supersaturation and the mixing mechanism might further depend on the dominant deformation behavior.

The synthesis of metastable phases during SPD deformation is further explained by the effective temperature model in literature [37], which was originally developed for materials under irradiation. During SPD, the material is driven into a state equivalent to a state at a specific increased (effective) temperature by the applied heavy deformation. Thus, metastable phase formations can occur in materials during SPD deformation.

Atomistic mixing between different metals to form supersaturated solid solutions, which are thermodynamically unstable and finally transform to an amorphous phase has already been reported in ref. [38,39]. The solid state amorphization process depend only on the degree of intense plastic deformation and can be accomplished within a much wider composition range compared to the frequently used solidification route [39]. Amorphization during HPT deformation has been observed in binary systems with a negative heat of mixing [40–43], but is not expected to occur in systems with a positive heat of mixing [38]. If alloys with glass forming tendency are HPT deformed, uniformly sized nanocrystals might also form in the amorphous matrix [38].

3. Structural and functional properties of HPT processed materials

The creation of novel HPT processed materials represents a promising pathway for obtaining materials with extraordinary mechanical and physical properties. For structural applications, mechanical properties like high strength combined with ductility are often expected. Furthermore, microstructural stability during enhanced temperatures is useful. These properties are also important for materials with interesting physical properties – i.e. magnetic, electric, radiation resistant - for functional applications. Nearly all materials discussed in the following section show excellent mechanical properties, which will not be separately discussed in this article. Special attention is instead given on first investigation results of selected physical properties. Additionally, the thermostability of the different HPT processed materials is addressed briefly.

3.1 Thermal stability

The thermal stability of NC supersaturated solid solutions, irrespective of whether single phase or multi-phase after HPT deformation, has been found to be significantly higher than expected for NC materials [38]. At low annealing temperatures, decomposition or phase separation is thermodynamically favored [38]. Thus, NC composite materials can be synthesized by phase separation during annealing before grain growth set in. In such a microstructure with isolated grains of each phase, which are further immiscible in each other, grain growth by boundary migration can be impeded. This grain growth retarding effect is already long known and also used in coarse-grained materials, i.e. in commercial dual phase steels, which consist of martensite islands in a ferrite matrix with phase ratio of 4:1.

HPT processed Cu-based metastable alloys from immiscible phase constituents have been extensively studied for their enhanced microstructural stability and strength. Investigated systems include Cu-Cr [19,44–46], Cu-W [47,48], Cu-Fe [46,49], Cu-Co [50,51] and Cu-Ta [52]. The thermostability of HPT deformed Cu-Cr alloys was examined in several studies [19,44–46]. In [19], a bulk, coarse grained Cu–Cr composite (43wt.% Cr, 57 wt% Cu) was processed by HPT. In the as-deformed material, a grain size of only 20 nm and the formation of a deformation-induced Cu supersaturated solid solutions were observed. Annealing at 450°C for 30 min resulted in decomposition and the formation of a stable Cu-Cr nanocomposite material with an even enhanced hardness compared to the as-deformed state. The same HPT deformed Cu-Cr material was later investigated in [44,45]. The examination of the microstructural evolution during HPT deformation and subsequent annealing by energy-dispersive synchrotron diffraction was the focus in ref. [44]. It could be shown that structural relaxation induced by phase separation during annealing and limitations in activating dislocation sources in the annealed material are mainly responsible for the observed enhanced hardness even though the domain (grain) size increases slightly during annealing. In [45], the thermal stability on the atomic scale was further investigated in-situ by spherical aberration-corrected transmission electron microscopy (TEM). It was shown that decomposition is accompanied by a decrease of the average interface width although grain growth is not yet observed.

In [47,48], a bulk, coarse-grained W-25% Cu alloy was deformed by HPT at room temperature (RT) resulting in the formation of a dual phase NC structure with partial solid solution formation. After deformation, the material was annealed at different temperatures. Up to 500°C, no grain growth was observed. After annealing at 720°C, limited grain growth up to 50 nm and complete phase separation took place. The obtained grain size in the nanocomposite after annealing was significantly lower compared to the undeformed W-Cu material.

Stacks of Cu and Ta foils were HPT deformed in [52] to obtain a high amount of mechanical mixing in the immiscible $Cu_{50}Ta_{50}$ system. After HPT deformation, the microstructure consisted of a Cu-16at%Ta solid solution, Ta-rich NC particles and pure Ta grains. This microstructure was thermally stable even after annealing at 1000°C. A fine structure was retained, which was attributed to the Cu-Ta solid solution and the NC Ta particles hindering grain boundary motion.

Instead of bulk dual-phase materials, Cu-Fe and Cu-Co powder mixtures with different initial composition were HPT deformed in [20,46,50,51]. In the HPT deformed Cu-Fe material, either complete solid solutions of Cu in Fe and Fe in Cu (for low Cu or Fe contents) or dual phase NC composite with partial supersaturation were obtained. A high thermal stability was observed in all deformed Cu-Fe materials. In [46], the structural evolution during annealing of a HPT deformed Cu-rich single phase solid solution was investigated in detail by in-situ annealing in the TEM and atom probe tomography (APT) investigations. Before deformation, the Fe-Cu powder mixtures used in [46] were compacted in air. Thus, oxygen contamination inevitably occurred. It was shown that oxide formation during subsequent annealing took place even before decomposition set in. These NC oxides were randomly distributed and contributed to the high thermal stability in the material.

A different phase separation process was observed in supersaturated Cu–Co solid solution with a slightly larger average grain size ( ~100 nm) in the as-deformed condition [51]. In the first stages of annealing, a NC scaled spinodal-type decomposition was observed in the grain interior. At the same time, pure alternating Co and Cu regions were formed near grain boundary regions. At higher annealing temperatures, an UFG microstructure consisting of Cu and Co was formed. Once phase separation was finished and the final microstructures were formed, they were stable even for very long annealing times of several days.

In summary, the amount of mechanical mixing and the achieved saturation grain size during HPT processing dictate the subsequent structural configuration of the immiscible phases after annealing. A high microstructural stability is achieved in all studies. If powder mixtures without protecting atmosphere are used as starting materials, oxygen contamination and oxide formation cannot be excluded. Oxide formation, however, further contributes to an enhanced thermal stability of the HPT processed materials.

## 3.2 Magnetic properties

The primary motivation for the investigation of NC magnetic materials results from the dramatic change in their hysteresis properties induced by the small grain size. First studies of magnetic characteristics of HPT processed materials were therefore conducted on single phase NC Cu and Co [53,54]. Nowadays, the magnetic properties of metastable materials synthesized by HPT deformation are studied as well. Tailored magnetic properties can be achieved by exploiting the effects of magnetic dilution in supersaturated solid solutions consisting of components with different magnetic properties (e.g. ferromagnetic and diamagnetic). The magnetic properties of the HPT processed metastable materials can be further modified by decomposition of these deformation-induced solid solutions to achieve unique distributions of ferromagnetic and diamagnetic phases.

In ref. [55,56], the size and distribution of Co and Fe particles in Cu based alloys were tuned by HPT deformation. After applying up to 10 rotations (Cu-22wt% Fe) and 25 rotations (Cu-10wt% Co), the ferromagnetic Co and Fe phases were fragmented into a size of ~100 nm and dissolution of some Co and Fe in the Cu matrix also took place, but no complete single phase solid solution was achieved up to the applied shear strain at room temperature. The coercive force in both alloys increased and the saturation magnetization decreased with increasing HPT strain, before both values reached saturation. Enhanced magnetoresistance was further observed (~2.5% at 77K and ~0.25% at RT for the Cu-Co alloy). Heat treatment at 723K for 1-10h led to a further increase of the measured magnetoresistance (~0.38% at RT). In ref. [57], Cu-Co alloys with only a small amount of Co (2.2 wt% and 4.9 wt% Co) have been studied. The coercivity after HPT-deformation was rather small for the Cu-2.2wt%Co alloy, which further showed superparamagnetic behavior because of Co particles with a size below the ferromagnetic limit (<10nm). The Cu-4.9wt% Co alloy showed a different behavior with a high coercivity due to larger Co particles.

Co-based alloys with low Cu content have also been investigated. In ref. [58], vacuum induction melted Cu-Co alloys (Co-5.6 wt% Cu and Co-13.6 wt% Cu) were processed by HPT deformation. Strong microstructural refinement in the Co phase (100 nm grain size) and Cu phase (precipitate size of 10 nm) was observed. As a result, the coercivity was drastically enhanced, while the saturation magnetization stayed almost unchanged. The supersaturated Cu-Co solid solution, which was reported to be present in the initial state after induction melting, completely decomposed during HPT deformation and a microstructure of almost pure Co grains and Cu precipitates was obtained.

Binary solid solutions in three different immiscible systems, consisting of ferromagnetic (either Co or Fe) and diamagnetic (either Cu or Ag) components, were synthesized by HPT in ref. [59,60]. In the Co-Cu alloys with medium compositions, deformation induced mixing of the immiscible elements and single phase ferromagnetic FCC solid solutions were obtained. With increasing Co content, microstructures with smaller grain sizes evolved as revealed by Transmission Kikuchi Diffraction (TKD) analysis (Fig. 1a). Additionally, the coercivity decreased with decreasing grain size (Fig. 1b), because exchange coupling of the NC grains resulted in randomization of the anisotropy. By annealing or HPT deformation at lower temperatures (liquid nitrogen temperature), the magnetic properties of the Co-Cu alloy with low Co content were further tailored [59]. In the Fe-Cu alloys, only samples with low Fe-content (<25wt.%) could successfully be deformed by HPT. Magnetic measurements revealed that Fe is clustered, but also partially diluted in Cu. Thus, a dual phase structure with a Cu-rich solid solution and a BCC Fe phase formed after HPT processing. Quite differently behaved the Fe-Ag system. No deformation-induced mixing was observed. The dual phase structure was further confirmed by magnetic measurements, which displayed no distortion of the Fe magnetic moment.

Alloys of the R-Fe-B system (R = Nd, Pr) are known as the most outstanding hard magnetic materials since their discovery in 1980s. Phase transformations and metastable phase formation generated during HPT deformation have also been successfully used in this material class with the objective to enhance the resulting magnetic properties. Already in the 1990s, Stolyarov et al. [61] investigated the effect of HPT deformation on the structure and phase formation in a $Pr_{20}Fe_{73.5}B_5Cu_{1.5}$ alloy. Besides grain refinement, which already led to an increase of the coercivity, non-equilibrium magnetically soft phases as well as an amorphous phase (from the main $Pr_2Fe_{14}B$ phase) were formed during deformation. Annealing after deformation restored the initial phases and the $Pr_2Fe_{14}B$ phase crystallized with a grain size of about 300 nm. A high coercivity of more than 1600 kA/m was achieved, which is higher compared to sintered and hot-rolled permanent magnets from the same type of alloy.

The influence of HPT deformation on microstructure, phase composition and coercivity of as-cast sub-, super- and stoichiometric $R_2Fe_{14}B$ (R = Nd, Pr)-based alloys was investigated in ref. [62]. Grain refinement to UFG structures and partial or complete decomposition of the $R_2Fe_{14}B$ phase took place after HPT processing in all investigated alloys. At the same time, a metastable phase – a BCC Fe-based solid solution rich in R-atoms - formed. The volume fraction of this metastable phase increased with increasing Fe content and was thus dependent on the initial alloy compositions. Annealing after HPT deformation resulted in complete or partial restoration of the initial phases, whereby the super-stoichiometric Pr-based alloy showed the highest coercivity.

Microstructure and magnetic properties of different as-cast and as-cast plus subsequently homogenized R–Fe–B alloys were studied after HPT deformation in ref.[61,63]. HPT deformation led to formation of UFG structures and even to the formation of an amorphous structure at extremely large strains. In the deformed as-cast alloys without homogenization treatment before HPT deformation, the crystalline α-Fe phase was additionally present. Annealing resulted in crystallization of the amorphous phase and the formation of stable magnetic phases. The microstructure consisted mainly of Pr-rich and $Pr_2Fe_{14}B$ phases in both alloys, which implies that α-Fe reacts with the excess Pr and transforms to the $Pr_2Fe_{14}B$ phase. A coercivity of 2232 kA/m was obtained for the Pr–Fe–B alloy, which was homogenized before deformation, and 1674 kA/m for the alloy, which was deformed in the as-cast condition. The lower coercivity in the latter case was related to a small fraction of remaining α-Fe phase in the microstructure after annealing.

In ref.[64], a Nd–Fe–B-based liquid-phase sintered alloy (composition of 66.5 wt% Fe, 22.1 wt% Nd, 9.4 wt% Dy, 1.0 wt% Co, 0.8 wt% B, 0.2 wt% Cu) was HPT deformed at RT. The starting alloy mainly consists of a $Nd_2Fe_{14}B$ phase. After deformation, grain refinement was observed and two different amorphous phases (Nd-rich and Fe-rich) with an embedded ferromagnetic $Nd_2Fe_{14}B$ phase were formed. A similar alloy (Fe–12.3 at% Nd–7.6 at% B) was investigated in ref. [65]. In contrast to ref. [64], an amorphous matrix with uniform composition, in which ferromagnetic $Nd_2Fe_{14}B$ nanocrystals were embedded, formed after HPT deformation of the as-cast alloy. To explain the formation of non-equilibrium phases, the authors argue, that HPT deformation at RT might be equivalent to an annealing treatment at an enhanced temperature. Therefore, an "effective" temperature can be determined according to the observed phases and their composition [37]. Thus, the "effective" temperature during HPT deformation in both investigated NdFeB-based alloys was estimated to be ~1140 °C and ~1170°C, respectively. Magnetic properties of the HPT deformed NdFeB-alloys have not been measured yet.

HPT deformation can also effect the phase evolution process of amorphous R-Fe-B-based alloys, which was used for the synthesis of hard magnetic nanocomposites by HPT deformation. In ref. [66], an over quenched, amorphous $Nd_9Fe_{85}B_6$ alloy was investigated. After deformation, decomposition of the amorphous structure and precipitation of nearly 40 wt% α-Fe with a size of about 10 nm was observed. The α-Fe precipitates served as nucleation sites for $Nd_2Fe_{14}B$ during a heat treatment at 600°C. Finally, a uniform and fine NC structure consisting of α-Fe and $Nd_2Fe_{14}B$ grains was obtained. Compared to the over-quenched and only annealed reference alloy, the coercivity and remanence increased by 23% and 16%, respectively. The effect of HPT on phase formation and magnetic properties of an amorphous $Nd_9Fe_{85}B_6$ alloy was also studied in ref. [67–69]. Precipitation of α-Fe nanocrystals, with an amount over 40%, occurred during deformation. It further lowered the amount of metastable phases formed after a subsequent annealing treatment compared to the rapidly quenched reference material. Furthermore, metastable phases with a high magnetization predominately formed in the HPT deformed material leading to a simultaneous enhancement of remanence and coercivity. A similar approach, but a different starting material, amorphous and partially amorphous $(Nd,Pr)_{10}Fe_{79}Co_3Nb_1B_7$ ribbons, were used in ref. [70,71] to obtain bulk α-Fe/$Nd_2Fe_{14}B$ nanocomposite magnets. Similar to ref. [67–69], the formation of metastable intermediate phases during subsequent annealing did not form in the deformed microstructure and only crystalline α-Fe and $Nd_2Fe_{14}B$ phases could be found by XRD investigations after annealing. Furthermore, a high fraction of α-Fe (>30%) was still present which resulted in an enhanced magnetization. At the same time, an increase in coercivity from 4.6 kOe (as-prepared amorphous ribbons only annealed) to 7.2 kOe (HPT processed and annealed material) was observed, which was attributed to an enhancement in domain wall pinning strength. Already in 2009, the reasons for the high amount of α-Fe in the HPT-processed amorphous R-Fe-B-based alloys was studied in ref. [72]. It was shown that vacancy-type defects formed in the deformed amorphous matrix, which were mainly surrounded with Fe atoms. Thus, the activation barrier for crystallization of the Fe phase was lowered, which leads to deformation-induced crystallization of the Fe phase in amorphous $Nd_9Fe_{85}B_6$ alloys.

HPT processing is a powerful tool for improvement of the magnetic properties of different magnetic materials using phase transformations or phase formations during deformation. The above studies show that an increase in coercivity or energy product is possible and also soft – or semi-hard magnetic materials can be successfully produced.

3.3 Electrical properties and superconductivity

Materials for electrical applications have to fulfill certain requirements, the most important are a high electrical conductivity, a high tensile strength and a good thermal stability. Ag, Cu, Au, and Al pure metals possess the highest electrical conductivities. Unfortunately, a combination of high strength and electrical conductivity is very hard to reach in these pure metallic materials. One possibility to increase the strength is alloying. However, introducing other elements into pure metals lowers its electrical conductivity due to electron scattering by solute atoms and by a minor contribution by precipitates. A NC grain structure is another option to achieve high strength, but the electrical conductivity is also reduced by electron scattering due to dislocations and grain boundaries in deformed materials.

A new approach combines grain refinement by HPT deformation with the formation of metastable phases or accelerated formation of nanosized second phases during processing. In ref. [73,74], the influence of HPT on strength and electrical conductivity in Cu-Cr alloy with very low (0.5%) [73] and high (up to 27%) [74] Cr contents has been investigated. The solubility of Cr in Cu at RT is nearly zero with a maximum solubility of about 0.9 at% at 1350K. Although not explicitly investigated, the dissolution of Cr and the formation of a supersaturated Cu solid solution was also expected for higher Cr-contents (9.85% and 27%) in ref. [74]. An increasing microhardness because of a decreasing grain size with increasing Cr content was further observed after HPT deformation. Additionally, the electrical conductivity decreased due to the higher amount of grain boundaries and Cr alloying atoms. The electrical conductivity of the Cu-alloy should, however, be less affected by Cr precipitates compared to Cr in solid solution. By annealing, supersaturated solid solution decomposition was achieved, which lead to an increased electrical conductivity. Simultaneously, a high hardness was preserved due to the restriction of considerable grain growth. Instead of HPT deformation at RT, the Cu-Cr alloy in ref. [73] was deformed at 300°C. A similar microstructure – 200 nm grain size of Cu with 10 nm Cr precipitates – was achieved by dynamic aging during HPT processing, which resulted in an enhanced electrical conductivity of 81%–85% IACS. Furthermore, the structure was stable up to 500°C.

A different Cu-based material was investigated in ref. [75]. Cu–NbC composites were synthesized starting from elemental Cu, Nb an C powders with different vol% using HPT deformation (5 GPa, 0.5 rpm, 20 turns). For the first time, the in situ formation of NbC by mechanical alloying was observed during HPT processing. Deformation induced phase formation took place in a significant shorter time than by ball milling, although the mechanism of NbC formation are proposed to be similar in both processes. As reason accelerated diffusion achieved by a reduced diffusion distances through the formation of very fine Nb particles and a reduced diffusion activation energy due to the high density of defects generated during HPT processing were proposed. The best property combination (high electrical conductivity and high tensile strength) was achieved in an HPT processed Cu–2 vol% NbC composites after annealing at 700 °C for 1 h.

Pure Al is often used as a material for conductors in the electrical industry due to its lower cost and lower density compared to Cu. HPT deformation and subsequent annealing was combined in ref. [76] to achieve high strength and high electrical conductivity in Al-Fe alloys (Al–2wt%Fe and Al–4wt%Fe). After HPT deformation, an UFG microstructure and partial dissolution of Fe in Al was achieved in both alloys [77]. From XRD peak shifts, 0.67wt% Fe and 0.99wt% Fe dissolved in the Al matrix after 75 rotations in the Al–2wt%Fe and Al–4wt%Fe, respectively. The electrical conductivity decreases with increasing HPT strain to final values of about 40% International Annealed Copper Standard (IACS). Annealing at 200°C resulted in precipitation of the dissolved Fe and the formation of well distributed, nanosized Fe particles. In the annealed state, an even higher strength compared to the as-deformed state as well as an enhanced conductivity (above 50% IACS) was achieved. A similar approach (HPT at RT with subsequent annealing or HPT deformation at 200°C) was used in Al-2wt%Fe alloys in ref. [78]. Both processing routes provided high strength, but a higher electrical conductivity (≥52%IACS) was achieved by HPT processing at elevated temperatures due to reduction of Fe solute atom concentration in the Al phase.

In ref. [79], the effect of HPT deformation on electrical conductivity of an as-cast Al–5.4wt% Ce-3.1wt% La alloy was investigated. Besides the development of an UFG microstructure (average grain size about 140 nm) with spherical intermetallic nanosized particles (Fig.2a), deformation induced supersaturated solid solution of the Ce and La in the Al matrix was further observed by HAADF-STEM imaging (Fig. 2b). Mixing was further confirmed by the change of the Al lattice parameter determined from XRD investigations. The concentration of the rare earth elements in Al was estimated to be about 0.1 to 0.2 at.%. These results show that mechanical alloying of Ce and La atoms in the Al matrix is possible even though they exhibit a very large atomic radius difference (about 30%). In the as-deformed state, the electrical conductivity decreased from 49.5% IACS (as-cast state) to 39.7% IACS. After annealing at 280°C for 1h, slight grain growth (average grain size about 200 nm), stable intermetallic particles and the formation of rare earth nanoclusters (~2 nm size) were observed. The electrical conductivity improved to 52.4% IACS with a high strength, both higher compared to the as-cast, undeformed state. In a subsequent study [80], the same rare earth elements (La + Ce) were used, but concentrations in the Al alloy were varied to define a composition, which provides an optimal combination of enhanced mechanical and electrical properties. Therefore, three different starting alloys with 0.9 wt% La and 1.6 wt% Ce, 2.9 wt% La and 1.6wt % Ce and 3.1 wt% La and 5.4 wt Ce wt% were produced by casting. Increasing rare earth element concentrations led to an enhanced strength, but reduced the electrical conductivity. For an optimal combination of electrical conductivity and mechanical strength, the total concentration of rare earth elements should not be above 4.5 wt%. Additionally, a subsequent annealing treatment between 250°C–280 °C for 1 h was shown to be beneficial for the electrical conductivity.

Nb–Ti alloys are widely used in another field of application, which are superconducting magnets. The influence of metastable phase and supersaturated solid solution formation on superconductive properties was investigated in a first study in ref. [81]. Nb-Ti powder blends with 47 wt% Ti were HPT deformed at RT and dissolution of Ti in Nb during deformation was observed. At large strains, a supersaturated NC β phase with BCC structure formed. During annealing, decomposition of the supersaturated phase, the formation of a lamellar structure and segregation of Nb at the grain and interphase boundaries was observed. The HPT processed alloy become superconductive at temperatures below 9 K. It was further shown that the transition temperature for superconductivity was lowered with increasing shear strain, but increased with subsequent annealing because of the decomposition process (Fig. 2c).

These first studies show that HPT processing (or more generally SPD) is a promising pathway to achieve materials with a good combination of electrical and mechanical properties.

3.4 Synthesis of porous materials and irradiation resistant materials

Applications of porous materials are lightweight structures, sensors or actuators, heat exchangers, dampeners or radiation tolerant materials, which require properties like low specific weight, high surface-to-volume ratio, excellent thermal and electrical conductivity, high energy absorption or large interface densities. Porous materials with ligament sizes in the UFG to NC regime especially have a high application potential. By varying the length-scale of the ligaments, the properties, e.g. the yield strength, can be adjusted. Conventional manufacturing methods are, for example, template-based fabrication or chemical dealloying of rapidly solidified alloys [82,83]. Recently, another procedure to obtain UFG and NC porous metallic materials based on HPT powder consolidation and deformation process has been developed [84–86]. Different combinations of immiscible powder mixtures (binary Cu/Fe [85] and Au/Fe [86] or ternary Cu-Fe-Ag [84] systems) were used as starting material resulting either in mechanically alloyed single phase or dual-phase NC microstructures. The bulk mechanically alloyed materials were then heat treated to reach phase separation or reduce the amount of forced mixing and/or to adjust the grain size of the respective phases. From these bulk materials, porous Cu, Au or Cu-Ag materials were then created by dealloying using selective etching with HCl [85,86] or potentiostatic dealloying [84].

In case of the porous Au material, a porosity of ~50% and ligaments with an average diameter of about 100 nm with on average ~ 70 nm diameter small grains was achieved [86]. For the porous Cu material, similar porosity levels and a slightly larger ligament size of about 200 nm was obtained [85]. Similar ligament sizes were observed for the Cu-Ag porous material [84]. As an example, the different microstructural states during the fabrication process of a $Cu_{50}Fe_{25}Ag_{25}$ material - initial state, as-deformed state, annealed, de-alloyed - leading to a dual-phase nanoporous material is illustrated in Fig. 3. In the SEM image of the initial state, the different phases are easily distinguished by their phase contrast (Fig.3a). After HPT deformation, a metastable single-phase was obtained (Fig.3b). After two different heat treatments, partial or complete phase-separated states with NC or UFG microstructures were obtained (Fig.3c-d). After selective potentiostatic dealloying of the Fe phase, a NC porous Cu-Ag material was achieved (Fig.3e).

To probe the mechanical properties and deformation behaviour of the nanoporous materials, nanoindentation was performed in ref. [84–86]. The Cu- and Au porous materials were further tested at different temperatures up to 300°C to evaluate the thermo-mechanical properties and structural stability. The hardness of the Au porous materials significantly increased after annealing, which was attributed to a significant reduction of mobile dislocations that left the material at free surfaces. Increasing hardness with increasing testing temperature was also observed for the Cu porous material. In this case, however, oxidization during high-temperature nanoindentation was observed and related to the measured mechanical properties. To study radiation effects and resistance, the Cu-Ag porous material was additionally tested before and after an irradiation treatment with 1 MeV protons to 1 dpa of damage at near RT. No notable difference in hardness before and after irradiation showed the proof-of-principle of a radiant-tolerant multi-phase UFG or NC porous material.

Another promising system for radiation tolerant properties are Cu-Nb nanocomposites. In Cu-Nb multilayer nanocomposites synthesized by different methods, radiation damage resistance in combination with high hardness has already been proven (i.e.[87–89] ). In [90], UFG Cu-Nb composites were synthesized by HPT processing and subsequent annealing. As starting material, Cu and Nb powders were mixed together with a composition of 50 wt% Cu and 50 wt% Nb. In the as-deformed state, a bulk NC material was obtained. After annealing at 500°C, an UFG Cu-Nb composites with grain sizes between 100-200 nm were obtained. Such UFG Cu-Nb composites might be used as material for applications in harsh radiation environments in future, but radiation tolerant properties have not been tested yet.

## 3.5 Synthesis of hydrogen storage materials

Hydrogen is regularly proposed as innovative energy carrier, but it remains a challenging task to design suitable and safe high-density hydrogen storage systems. Different types of hydrogen storage, like pressurized gas tanks, storage as liquid hydrogen or chemical storage as hydrocarbons, already exist. Another promising way of hydrogen storage is in reversible, solid state, light metal hydrides (such as Mg, $MgNi_2$, $LaNi_5$ or TiFe), but they exhibit slow kinetics in the charging/discharging processes and high hydrogen desorption temperatures. The kinetics can, however, be radically enhanced by reducing the particle size and/or by introducing lattice defects, which enhance diffusion and can act as pathways for hydrogen transport. In 2010, the microstructure and hydrogen sorption properties of Mg, $MgH_2$, and $MgH_2$–Fe powder mixtures after HPT deformation were investigated [91]. Besides grain refinement with average grains sizes of 20 nm, also the formation of a metastable $\gamma$-$MgH_2$ phase and enhanced hydrogen sorption properties were observed in the bulk samples. A Japanese research group further determined the potential of the HPT powder consolidation process to synthesize novel materials for possible hydrogen storage application at RT using deformation-induced mixing [92]. In particular, different Mg-based alloys have been examined so far and the phase formations have been further verified by first-principle calculations. Over 20 different elemental powders such as Ti, Zr, Al and Zn were mixed with Mg powders and deformed by HPT at RT [93]. Various nanostructured intermetallics, but also new metastable intermetallics, alloys or amorphous phases were synthesized.

The Mg–Ti binary system, for example, is immiscible even in liquid form, but both elements react with hydrogen forming $MgH_2$ and $TiH_2$ hydrides. With HPT processing, metastable phases (BCC, FCC and two HCP structures) were formed in the NC material as shown by XRD and TEM investigations in ref. [94]. The formation of the metastable phases was explained either by deformation induced atomic-scale mixing (for both HCP phases) or the effect of grain size on phase stability (for the BCC and FCC phases). After hydrogenation, $MgH_2$ and $TiH_2$, but no Mg-Ti hydrides were formed although first-principles calculations showed that this hydrogenation reaction should occur thermodynamically. Thus, decomposition of the metastable Mg–Ti binary phases to pure Mg and Ti is kinetically faster than hydrogen adsorption in the form of ternary Mg–Ti hydrides with the cubic structure.

Mg and Zr are also immiscible elements, which means they do not form any binary phases in equilibrium. Similar to Mg, Zr can store hydrogen in the form of $ZrH_2$. Powder mixtures of Mg - 50 at% Zr were HPT deformed and several new metastable phases (nanostructured HCP, nano-twinned FCC, BCC or ordered BCC-based phases) were obtained [95]. At large shear strains, the deformed Mg-Zr material consisted mainly of a HCP Mg-Zr solid solution and small amounts of BCC and FCC phases. The BCC phase almost vanished at the highest amount of strain, indicating that it is only an intermediate phase during phase transformation to the HCP Mg-Zr phase. TEM high-resolution lattice images and high-resolution EDS maps further showed many Mg-based nanoclusters in the microstructure (Fig.4a). Contrary to the investigated Mg-Ti [94], the non-equilibrium Mg-Zr phase did not decompose until temperatures of 773K. The HPT processed Mg-Zr material further exhibited reversible hydrogen storage capability with an absorption of ~1 wt.% of hydrogen mainly in the Mg-based nanoclusters (Fig.4b). The absorbed hydrogen further fully desorbed under air or argon atmospheres at RT.

No new phase formation, but amorphization was observed, for example, in the Mg–Al system [93]. Furthermore, the desorption temperature for hydrogen decreased and an increase in hydrogen storage capacity was observed. The hydrogen storage capacity was, however, lower than for HPT deformed pure Mg.

Not only binary Mg-alloys have been investigated so far, also ternary Mg-based systems have been processed by HPT [96,97]. Compositions in the Mg−V−Sn, Mg−V−Pd, Mg−V−Ni, Mg−Ni−Sn, and Mg−Ni−Pd were first selected based on first-principles calculations. Initial materials were either powder mixtures (Mg−V with Sn, Pd or Ni) or as-casted ingots (Mg−Ni−Sn or–Pd), which were subsequently HPT deformed to ultra-high strains. XRD investigations proved the formation of new metastable phases (B2-type structure in Mg-V-Sn, B2-type and BCC structure in Mg-V-Pd, single phase BCC structure in Mg-V-Ni) besides thermodynamically stable $Mg_2Sn$ and $Mg_2Ni$ intermetallics. In the Mg-V-Ni system, complete mixing of elements and structural saturation was observed. In Mg-V-Sn and Mg-V-Ni, structural inhomogeneities are visible even after extremely large shear strains. In the Mg-Ni-Sn and Mg-Ni-Pd systems, new phases are formed as well. Uniform atomic mixing of initial three intermetallics to a new single phase with a BCC-based partly ordered CsCl-type structure was confirmed by APT investigations in Mg-Ni-Pd [97]. The novel HPT-synthesized alloy exhibited high phase stability and reversible hydrogenation and dehydrogenation properties at RT. In contrast, the Mg-Ni-Sn transformed from a microstructure containing three intermetallics in the as-cast state to a mainly amorphous structure with some nanograins. Unfortunately, no results of hydrogen related properties are published so far for this system.

New Mg-V-Cr alloys were further synthesized by a combination of prior ball milling and subsequent HPT of mixtures of $MgH_2$, V and Cr powders with MgV, $Mg_2VCr$, $MgV_2Cr$ and MgVCr compositions [98]. Novel single phase BCC phases are formed for the compositions $Mg_2VCr$ and MgVCr, whereas in MgV and $MgV_2Cr$ a new BCC phase was formed in addition to the initial phases. Best properties regarding structural stability and hydrogen storage at RT was achieved in the MgVCr alloy.

In ref. [99], a Ti-50 at.% V alloy with a supersaturated BCC structure was synthesized at RT from Ti and V powders using HPT deformation. Although hydrogen diffusion and metal-to-hydride phase transformation could be improved, the rate of hydrogen dissociation was still slow.

It can be summarized that the production of innovative hydrogen storage materials with unique compositions and microstructures is possible using HPT deformation.

## 4. New material combinations
### 4.1 Amorphous materials and bulk metallic glass composites

Amorphous materials or bulk metallic glasses have several advantages as compared with crystalline metals, but have also some major drawbacks such as their brittleness and poor ductility in tensile testing. One way to influence the properties is to produce bulk metallic glass composites, which contain an additional amorphous or crystalline phase as inhomogeneity [100]. Usual processing routes are, for example, partial crystallization of amorphous samples, casting metallic glasses with crystalline parts such as fibers, or spark plasma sintering [101–107].

A new processing approach is to generate bulk metallic glass composites by HPT deformation starting from powders, in which atomic mixing or mechanical alloying between the constituent phases is also observed. As starting materials, mixtures of amorphous/crystalline or amorphous/amorphous powders are used. Up to now, Zr-metallic glass powders in combination with crystalline Al or Cu powders and mixture of Zr- and Ni-based metallic glass powders have been investigated. In ref. [108], 60 vol.% Zr-based metallic glass and 40 vol% Al crystalline powders were spark plasma sintered to obtain a pre-compacted initial material. Investigations of the microstructural evolution during HPT deformation have shown that most of the deformation was localized in the Al matrix. The metallic glass phase was, however, also deformed and became fragmented, which led to microstructural refinement. Inhomogeneous deformation and shear band formation was further observed. Further HPT deformation led to full amorphization and intensive mechanical mixing. For the highest strains (larger than 400), a featureless, homogeneous structure was observed, which also correlates to the hardness maximum of 700HV. Shear bands occurring with a high density during HPT deformation were assumed to be the main mechanism for the observed mixing. It was shown by a simple geometrical model that few atomic jumps within shear bands could be enough to fully mix and amorphize the composite, if the level of deformation is large enough. However, chemical gradients with a typical length scale in a range of 10–20 nm were still present in the microstructure even at high strains. The very high hardness of the as-processed composite, which is 40 % higher than the hardness of the original Zr-based metallic glass, was further attributed to these chemical gradients.

Zr-based metallic glass powders with four different compositions (20, 40, 60, 80 wt% Cu) were mixed with crystalline Cu powders, then directly consolidated and deformed by HPT in ref. [109]. Although the Cu phase carried the most part of the deformation and no ideal co-deformation occurs, phase refinement of both phases into a lamellar structure was observed (Fig.5a-f). The structure refined further with on-going deformation and in saturation at very high strains, single phase bulk metallic glasses were obtained for all composition except the one containing 80 wt% Cu. In this case, a part of the Zr-metallic glass phase was dissolved in the Cu phase forming a supersaturated solid solution, which remains crystalline. Furthermore, it was observed that the strain necessary to reach saturation increases significantly with increasing content of the crystalline Cu phase. The soft Cu phase carried most of the deformation; thus the amorphous particles are not forced to deform as strongly as compositions with lower Cu content. Therefore, fully mixing might be prevented as well.

Mechanical properties of the HPT deformed bulk Zr-metallic glass-based materials were investigated with hardness measurements. The compositions with 20, 40 and 60 wt% Cu exhibited a hardness that is lower than a reference Zr-metallic glass at low strain. In the fully mixed amorphous state, the hardness was, however, higher than the pure Zr-metallic glass reference. By contrast, the composite containing 80 wt% Cu showed on the one hand lower hardening with increasing strain and on the other hand the hardness of the reference Zr-metallic glass cannot be obtained even at the highest investigated strain.

A similar combination of starting materials ($Zr_{55}Cu_{30}Ni_5Al_{10}$ - metallic glass phase in combination with crystalline Cu) was used in [110], but instead of powders plates with semi-circular shape and a volume ratio of 1:1 were put together and subjected to HPT. SEM and TEM investigations showed that lamellar nano-composites were produced at the highest strain (50 rotations). In contrast to [109], no mixing or chemical reaction between the crystalline/amorphous phases or crystallization of the amorphous phase was observed by XRD. The reason might be the significant smaller amount of applied strain (50 rotations) compared to ref. [109] (500 rotations). Additionally, a large scatter in hardness was determined after deformation. The average hardness was also slightly smaller compared to the as-cast $Zr_{55}Cu_{30}Ni_5Al_{10}$ reference material.

In ref. [111], two different metallic glass powders, $Zr_{57}Cu_{20}Al_{10}Ni_8Ti_5$ and $Ni_{53}Nb_{20}Ti_{10}Zn_8Co_6Cu_3$, were consolidated and deformed by HPT to synthesize amorphous dual phase composites. During HPT deformation, deformation of both amorphous phases took place and a lamellar structure similar as in the amorphous/crystalline composites evolved. The structure refines down to a few nm during deformation. Shear band formation was observed as well and at very high shear strains even mixing of the two amorphous materials towards a single phase bulk metallic glass occurred. The newly formed amorphous phase has a hardness between the initial metallic glasses, but higher than the rule of mixture predicts.

In a slightly different approach, single amorphous powders are deformed and deformation-induced crystallization resulted in bulk metallic glass composites as well. In this way, NC bulk $Al_{90}Fe_5Nd_5$ partially amorphous composites with Al nanocrystallites with a size between 5 and 25 nm were prepared "in-situ" by HPT deformation of fully amorphous gas atomized powders [112]. Instead of amorphous powders, ball milled $Fe_{77}Al_{2.14}14Ga_{0.86}P_{8.4}C_5B_4Si_{2.6}$ amorphous ribbons were used in ref. [113]. HPT deformation led to the formation of dispersed $Fe_2B$ or α-Fe nanocrystals inside the amorphous matrix. The bulk metallic glass composites exhibited enhanced Curie temperature and hardness with respect to the not deformed ribbon material.

The HPT process permits the production of bulk metallic glass composites. Even novel metallic glass phases with different chemical composition can be synthesized. By varying the applied strain, phase dimensions can be systematically varied and deformation-induced mixing is observed. However, enormous amounts of strain are necessary to achieve full-mixing. With this process, bulk metallic glass composites containing metallic materials with new types of second phase - for example, Cu, which would be impossible to be obtained by classical solidification methods due to the easy melting of Cu- can be synthesized. Additionally, a much wider composition range not accessible by the classical routes, is possible. As an example, the wide range of chemical compositions achieved in ref. [109] are shown in Fig.5g.

## 4.2 High-entropy alloys

High-entropy alloys, which are complex alloys consisting of five or more principal elements, have received considerable attention in the material science community in the last few years [114,115]. High-entropy alloys with micrometer sized grains are usually processed by melting and casting of the pure elements followed by rolling and recrystallization [116]. HPT deformation of as-cast or arc-melted coarse grained high-entropy alloys has been reported to lead to grain sizes in the NC regime [117–120] leading to extraordinary high strength and structural stability. Despite their excellent properties, it is a rather long process – casting and subsequent deformation- to achieve the desired nanostructure. As alternative, processing of bulk NC high-entropy alloys directly from the constituent powders using deformation-induced alloying by HPT as effective and time-saving method was recently proposed in ref. [121,122]. Powder mixtures of equiatomic Co, Cr, Fe, Mn and Ni were pre-compacted and subsequently HPT deformed up to 100 rotations with a pressure of 5 GPa. As revealed by XRD, the microstructure first evolved into a (FCC+ BCC) based solid solution after 10 rotations. After 100 rotations, a FCC single phase solid solution with an increased lattice spacing for the FCC lattice was obtained. TEM investigations showed a NC microstructure with small amount of chromium oxide precipitates (size of 7–10 nm). APT analysis further proved alloying and a homogeneous distribution of the major constituting elements on the nanometer scale (Fig.6). The hardness of the as-deformed alloy showed high values of about 6700 MPa. The higher values compared to HPT deformed as-cast or arc-melted CoCrFeMnNi alloys (4900–5380 MPa [117–120]) with similar grain size are explained by additional precipitation hardening of the nanometer sized chromium oxide particles. The observed mechanical alloying was explained by accelerated atomic diffusivity and defect introduction under HPT conditions in combination with intensive mass transfer across interfaces by "superdiffusive" shear-induced mixing [28].

In summary, the potential of HPT processing as innovative and effective route for synthetization of high-entropy alloys is demonstrated, with many more options in choosing the range of compositions in future. However, we assume that processing of single phase high-entropy alloys, which are desirable to achieve most outstanding properties, by deformation-induced mixing might be only possible in a limited temperature range and for compositions close to those which have been found to crystallize as single-phase solid solutions with conventional processing routes as well.

4.3 Innovative starting materials

An important issue for deformation-induced mixing is the homogeneity of the deformation process, which is also influenced by the structural homogeneity of the starting material. Often powder mixtures are used as starting materials and inhomogeneity is a well-known problem. It is especially likely to occur when the various components of the powder mixture differ strongly in size and density. One solution might be laborious techniques of mixing, another the use of coated powder particles as initial starting material, which means that any desired combinations of thin metal coatings and substrate metal might be attained.

In a first feasibility study [123], Fe powders coated by a Cu layer of about 1-2 µm thickness were prepared by immersion deposition and subsequently inductively hot-pressed to obtain a pre-compacted starting material. For comparison, a mixture of elemental Fe and Cu powders with similar powder particle sizes and overall composition was pre-compacted in the same way as the coated powders. Huge differences were already observed after this initial inductive hot-pressing step. The initial microstructure of the pre-compacted coated powders consists of Fe cores separated by a Cu layer network. Furthermore, the compacted material was very homogeneous on a macroscopic scale (Fig.7a). On the contrary, the microstructure of the pre-compacted elemental powder mixture was chemically inhomogeneous (Fig. 7b) with concentrations scattering on a length scale that is even comparable to the size of small HPT samples (diameter of 8-10 mm). A quite coarse initial Cu and Fe microstructure with each a size up to several hundred microns was also observed due to particle clustering. After HPT deformation, this initial structure leads to inhomogeneous deformation, strain concentrations and non-symmetric hardness profiles. On the very contrary, the initial higher homogeneity of the coated powder sample results in a more uniform deformation during HPT processing, a higher hardness even at lower strain and a reduced amount of strain necessary to reach the saturation state. Furthermore, the starting material influences the mechanical mixing process as well. A larger amount of Fe was dissolved in the Cu phase, and vice versa, for the severely deformed coated powder in this saturation state as revealed by XRD measurements. Possible reasons are the smaller grain size and higher densities of stored dislocations, which are both stabilized by a higher amount of impurities. As shown by XRD, the coated powder contained a significant amount of oxides which is higher than for the powder mixtures.

Another approach to prevent powder inhomogeneity would be the synthesis of pre-alloyed powders. With inert gas condensation, for example, the production of pure metals, alloys and oxides with average NC grain sizes, high purity and uniform nanostructures is possible. Drawbacks are, however, the low production rates and that only a limited number of metals and alloys that can be vaporized. Another option would be the use of gas atomized powders as starting material. Up to now there exists, however, only one study using gas atomized powders in combination with HPT deformation [112]. A vision of the future would be the production of materials by design using the powder synthesis route, where multiple phases in the form of powder particles might be combined and consolidated by HPT deformation.

5. Summary and Outlook

HPT is a SPD technique, which, in addition to grain refinement, can be used to synthesize bulk metastable materials and novel nanocomposites having UFG or NC structures. The UFG or NC grain size of these materials does not only result in advanced mechanical properties. More importantly, metastable phases and alloys with compositions beyond the equilibrium phase diagrams and with fewer thermodynamic restrictions can be designed. Most of the studies aim to develop innovative bulk materials with superior functional properties for future applications. The important points of this overview can be summarized as follows:

- For nearly all HPT synthesized bulk materials, ultrahigh strengths are reported. Furthermore, an enhanced thermostability compared to pure metals is observed.
- The magnetic properties of HPT processed materials can be tailored using phase transformations or phase formations. That applies both to hard magnetic and to soft magnetic materials.
- High strength materials with good electrical conductivity for electrical applications can be processed as well.
- First promising feasibility studies on the synthesis of porous materials, with efficient resistance to radiation, and irradiation resistant nanocomposites using HPT deformation have been conducted.

- Innovative materials for solid hydrogen storage application with enhanced kinetics have been produced using deformation-induced mixing and phase formations by HPT.
- The first papers on amorphous materials, bulk metallic glass composites and high-entropy alloys synthesized by HPT deformation, which show promising results for the future, as well as the use of innovative starting materials are presented.

Acknowledgments

Funding of this work has been provided from the European Research Council (ERC) under the European Union's Horizon 2020 research and innovation programme under ERC Grant Agreement No. 340185 USMS and No. 757333 SpdTuM.

# REFERENCES

[1] P.W. Bridgman, Phys. Rev. 48 (1935) 825–847.

[2] K. Edalati, Z. Horita, Mater. Sci. Eng. A 652 (2016) 325–352.

[3] V.. Stolyarov, Y.. Zhu, T.. Lowe, R.. Islamgaliev, R.. Valiev, Nanostructured Mater. 11 (1999) 947–954.

[4] R.Z. Valiev, Y. Estrin, Z. Horita, T.G. Langdon, M.J. Zehetbauer, Y. Zhu, JOM 68 (2016) 1216–1226.

[5] Y. Huang, T.G. Langdon, Mater. Today 16 (2013) 85–93.

[6] Y. Estrin, A. Vinogradov, Acta Mater. 61 (2013) 782–817.

[7] X. Sauvage, A. Chbihi, X. Quelennec, J. Phys. Conf. Ser. 240 (2010) 012003.

[8] B. Straumal, A. Korneva, P. Zięba, Arch. Civ. Mech. Eng. 14 (2014) 242–249.

[9] A.V. Korznikov, G. Tram, O. Dimitrov, G.F. Korznikova, S.R. Idrisova, Z. Pakiela, Acta Mater. 49 (2001) 663–671.

[10] S.D. Prokoshkin, I.Y. Khmelevskaya, S.V. Dobatkin, I.B. Trubitsyna, E.V. Tatyanin, V.V. Stolyarov, E.A. Prokofiev, Acta Mater. 53 (2005) 2703–2714.

[11] Z. Kovács, P. Henits, A.P. Zhilyaev, Á. Révész, Scr. Mater. 54 (2006) 1733–1737.

Captions List

Fig.1: TKD images of Co28wt.%-Cu (a), Co49wt.%-Cu (b), and Co67wt.%-Cu (c) in the as-deformed state. In (d), coercivity and saturation magnetization are plotted for the different Co compositions. The grey dotted line indicate the magnetic moment of FCC-Co [60].

Fig.2: (a) STEM-HAADF image showing the microstructure of the HPT deformed Al–5.4wt% Ce-3.1wt% La alloy with a small particle on a grain boundary. The EDS line profile analysis along the grain boundary (indicated by the red arrow) indicates Ce and La segregation. (b) High resolution HAADF STEM image in [001] zone axis of the Al matrix. The atomic column, which are significantly brighter that the average are suggested to contain at least one rare earth atom [79]. (c) Transition temperature for superconductivity Tc as function of shear strain and annealing time [81].

Fig.3: Different microstructural states during fabrication of Cu-Ag porous material by HPT [84]. (a) SEM micrograph of the initial Cu-Fe-Ag material. (b) Bright field TEM micrograph of the material after HPT deformation. (c) Bright field TEM micrograph after heat treatment at 400°C for 1h. (d) Bright field TEM micrograph after heat treatment at 600°C for 1h. (e) SEM

micrograph of the Cu-Ag porous material.

Fig.4: (a) High-resolution HAADF lattice image and corresponding EDS mappings of Mg-based nanoclusters at the highest amount of strain. (b) Hydrogenation kinetics after HPT deformation [95].

Fig.5 SEM images of a HPT processed Zr-40wt%Cu metallic glass composite at different shear strains: (a) shear strain 30, (b) 60, (c) 120, (d) 250, (e) 570 and (f) 850. The magnification is the same in all images. (g) Bulk Zr metallic glass composites obtained by HPT processing [109], in which the chemical composition is adjusted by dissolving Cu into the Zr-metallic glass ($Zr_{57}Cu_{20}Al_{10}Ni_8Ti_5$). The extensive range of possible compositions is marked by the red backgrounded field. In X, all elements of the metallic glass except Cu and Zr are summed up. The colored data points correspond to chemical composition of bulk metallic glasses from literature (please refer to [109] for details).

Fig.6: APT reconstruction showing the elemental distribution of only 1% of all Co, Cr, Fe, Mn and Ni ions, but all detected CrO and O ions in the as-deformed state [121].

Fig.7 Light micrograph images of cross-sections of the hot-pressed coated powder (a) and the hot-pressed powder mixture (b). The overlay illustrates EDX

measurement areas with the color indicating the measured Cu contents in wt.% in each area [123].

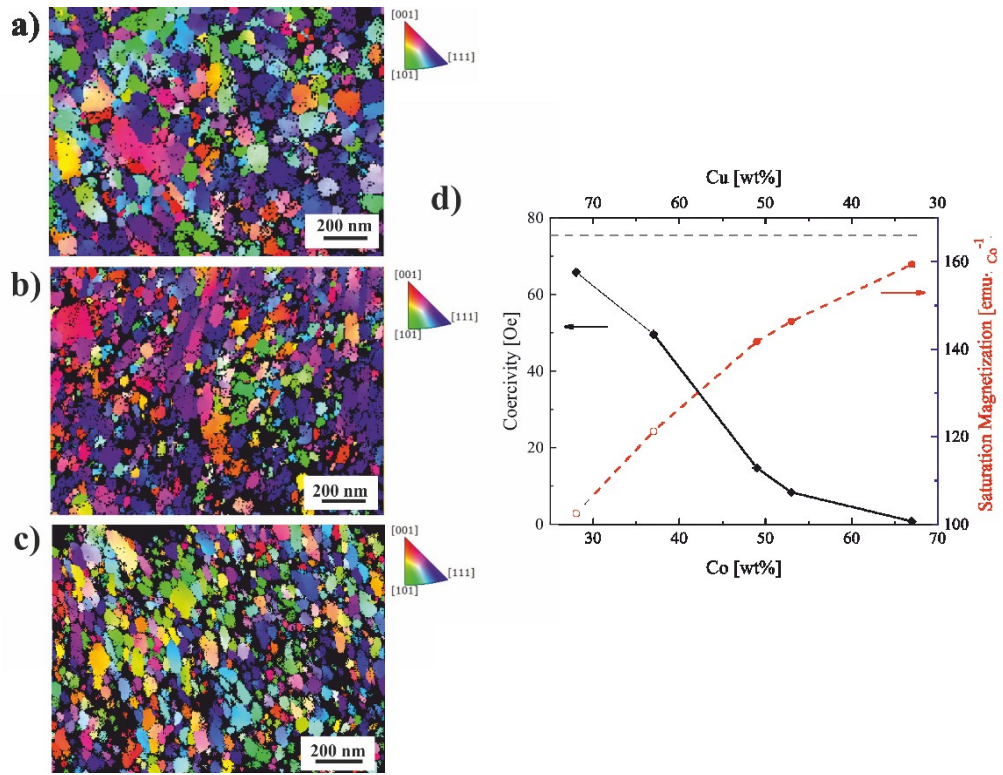

Fig.1 TKD images of Co28wt.%-Cu (a), Co49wt.%-Cu (b), and Co67wt.%-Cu (c) in the as-deformed state. In (d), coercivity and saturation magnetization are plotted for the different Co compositions. The grey dotted line indicate the magnetic moment of FCC-Co [60].

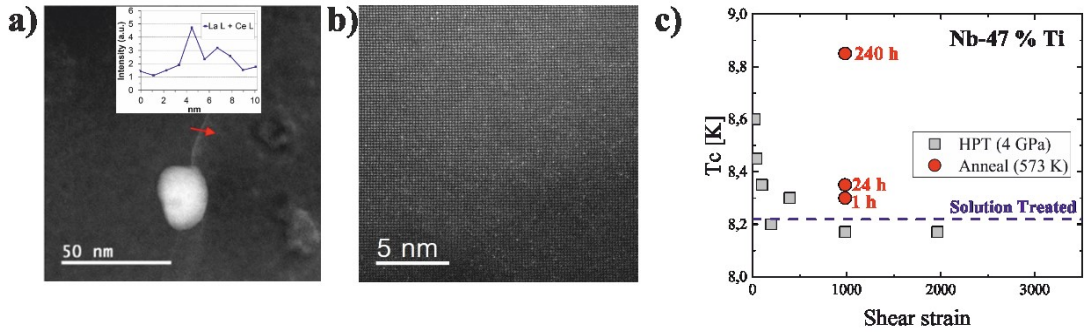

Fig.2 (a) STEM-HAADF image showing the microstructure of the HPT deformed Al–5.4wt% Ce-3.1wt% La alloy with a small particle on a grain boundary. The EDS line profile analysis along the grain boundary (indicated by the red arrow) indicates Ce and La segregation. (b) High resolution HAADF STEM image in [001] zone axis of the Al matrix. The atomic column, which are significantly brighter that the average are suggested to contain at least one rare earth atom [79]. (c) Transition temperature for superconductivity Tc as function of shear strain and annealing time [81].

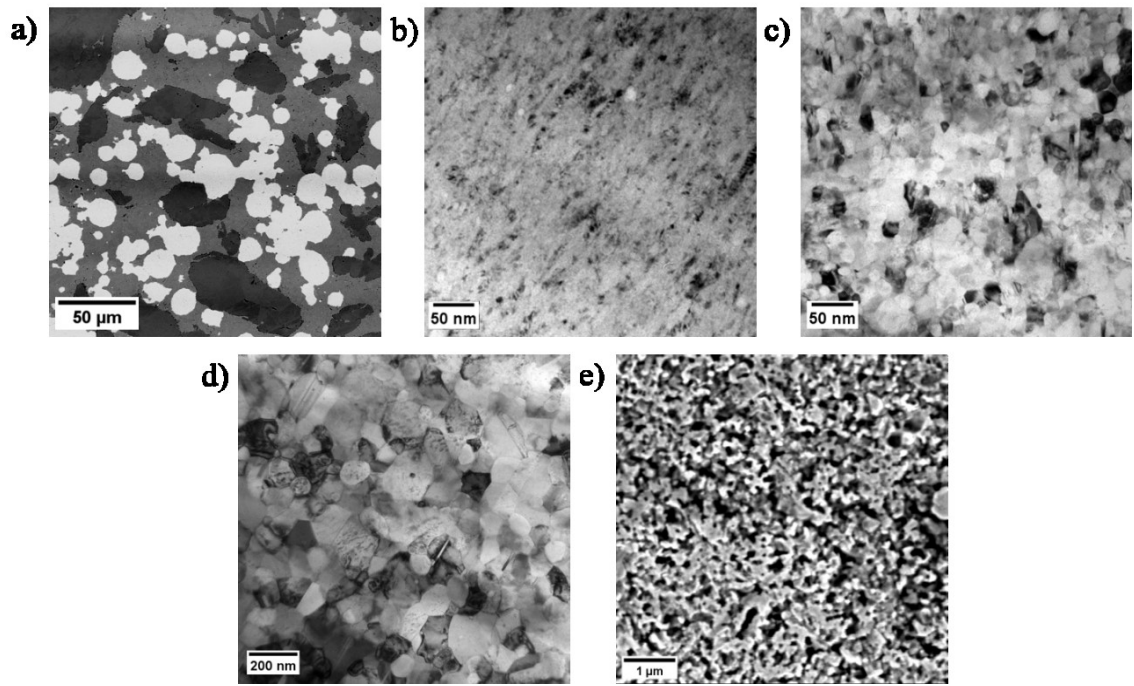

Fig.3 Different microstructural states during fabrication of Cu-Ag porous material by HPT [84]. (a) SEM micrograph of the initial Cu-Fe-Ag material. (b) Bright field TEM micrograph of the material after HPT deformation. (c) Bright field TEM micrograph after heat treatment at 400°C for 1h. (d) Bright field TEM micrograph after heat treatment at 600°C for 1h. (e) SEM micrograph of the Cu-Ag porous material.

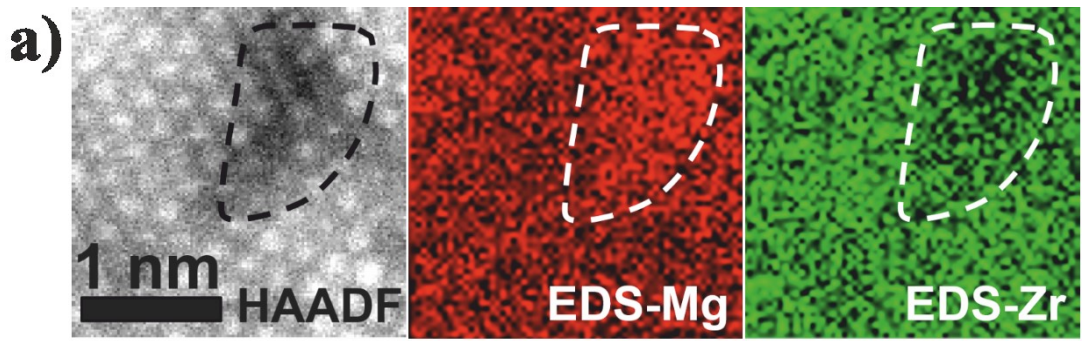

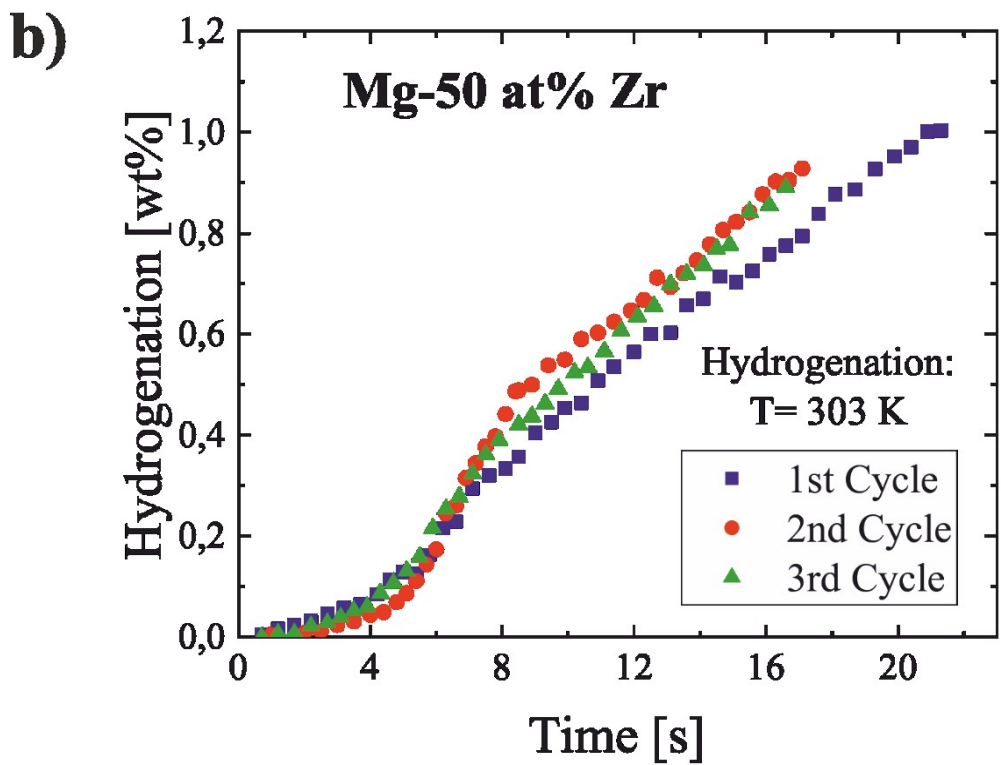

Fig.4 (a) High-resolution HAADF lattice image and corresponding EDS mappings of Mg-based nanoclusters at the highest amount of strain. (b) Hydrogenation kinetics after HPT deformation [95].

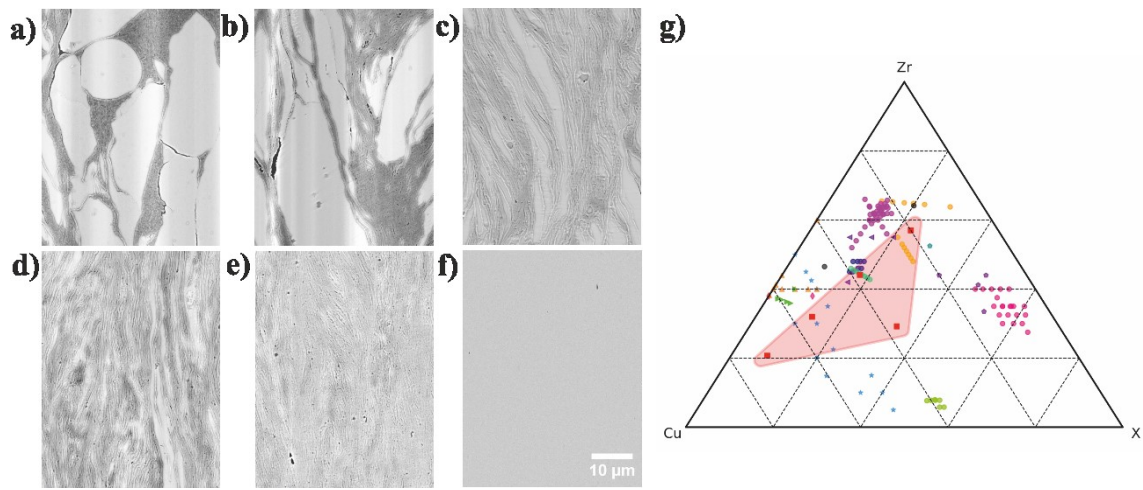

Fig.5 SEM images of a HPT processed Zr-40wt%Cu metallic glass composite at different shear strains: (a) shear strain 30, (b) 60, (c) 120, (d) 250, (e) 570 and (f) 850. The magnification is the same in all images. (g) Bulk Zr metallic glass composites obtained by HPT processing [109], in which the chemical composition is adjusted by dissolving Cu into the Zr-metallic glass ($Zr_{57}Cu_{20}Al_{10}Ni_8Ti_5$). The extensive range of possible compositions is marked by the red backgrounded field. In X, all elements of the metallic glass except Cu and Zr are summed up. The colored data points correspond to chemical composition of bulk metallic glasses from literature (please refer to [109] for details).

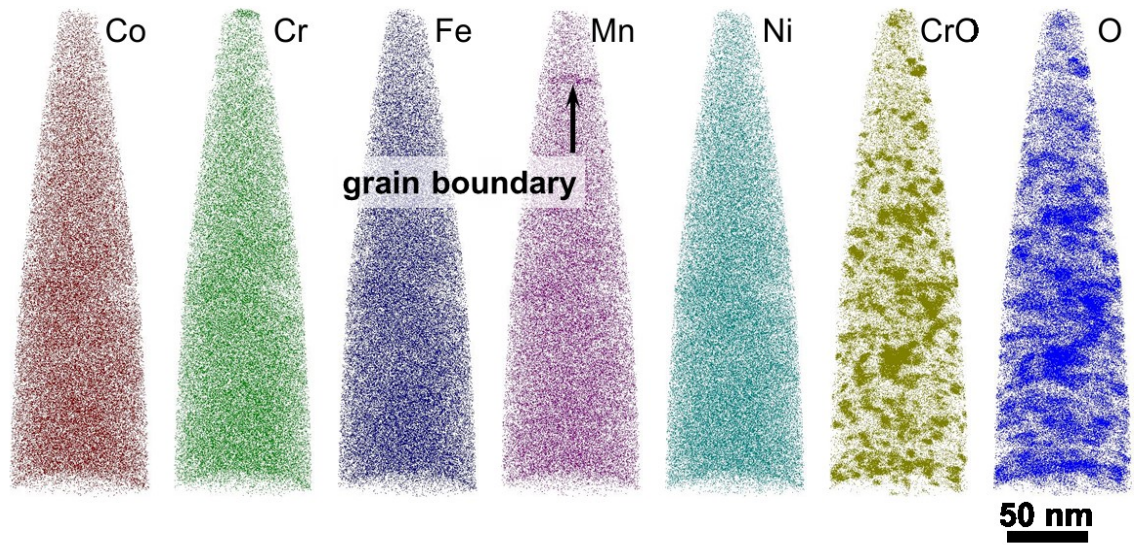

Fig.6 APT reconstruction showing the elemental distribution of only 1% of all Co, Cr, Fe, Mn and Ni ions, but all detected CrO and O ions in the as-deformed state [121].

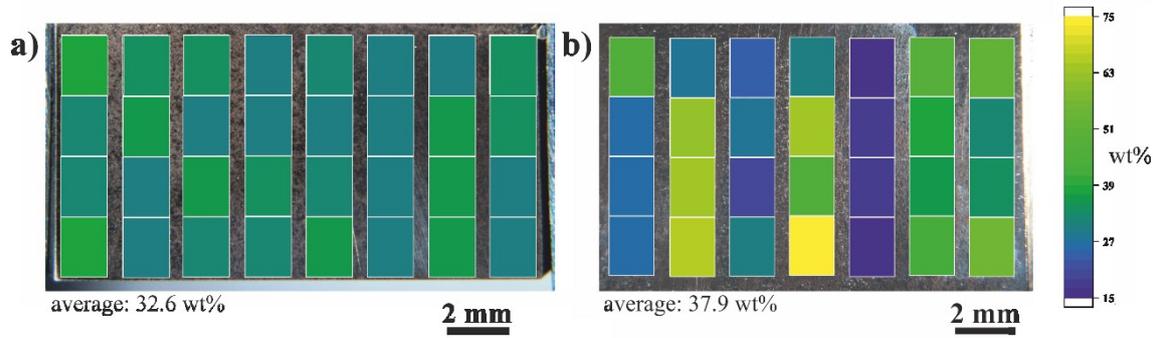

Fig.7 Light micrograph images of cross-sections of the hot-pressed coated powder (a) and the hot-pressed powder mixture (b). The overlay illustrates EDX measurement areas with the color indicating the measured Cu contents in wt.% in each area [123].

Graphical abstract

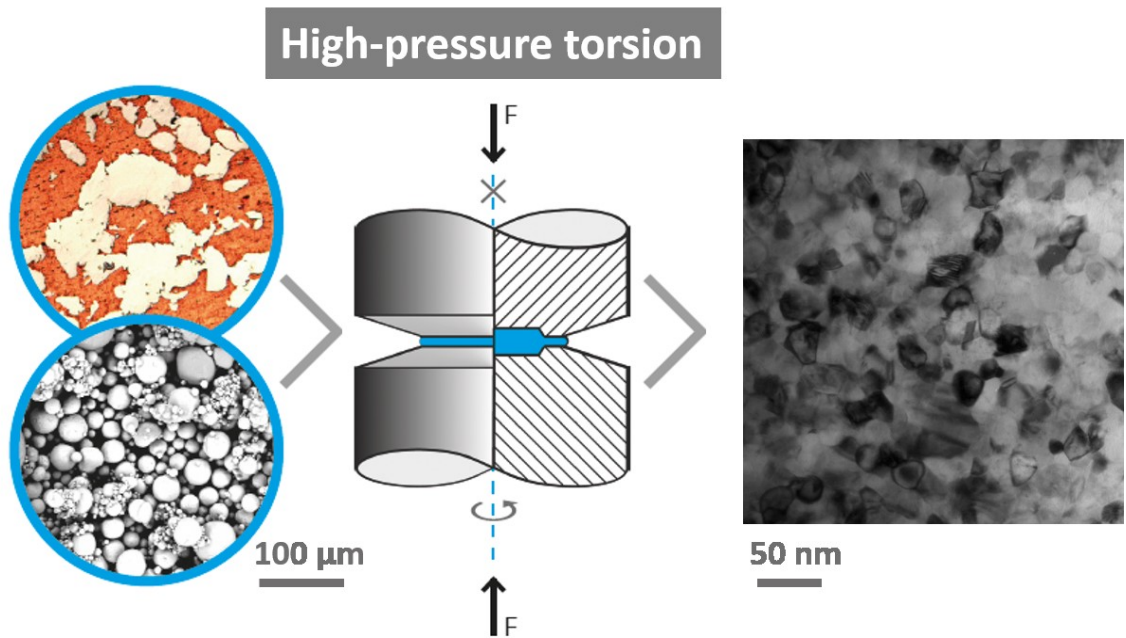